\documentclass{natureTEST}

\usepackage{graphicx}
\usepackage{bm}
\usepackage{amsmath,amssymb}
\usepackage{xspace}

\usepackage{textcomp}
\usepackage[T1]{fontenc} 

\usepackage{color}
\definecolor{bl}{rgb}{0, .1, .6}
\definecolor{rd}{rgb}{1,0,.2}

\title{Observing the Rosensweig instability of a quantum ferrofluid}
\author{Holger Kadau$^{1}$, Matthias Schmitt$^{1}$, Matthias Wenzel$^{1}$, Clarissa Wink$^{1}$, Thomas Maier$^{1}$, Igor Ferrier-Barbut$^{1}$ \& Tilman Pfau$^{1}$}

\begin{document}

\maketitle

\begin{affiliations}
	\item 5. Physikalisches Institut and Center for Integrated Quantum Science and Technology, Universit\"{a}t Stuttgart, Pfaffenwaldring 57, 70569 Stuttgart, Germany
\end{affiliations}

\begin{abstract}
Ferrofluids show unusual hydrodynamic effects due to the magnetic nature of their constituents. For increasing magnetization a classical ferrofluid undergoes a Rosensweig insta\-bi\-lity\cite{RosensweigFerrohydrodynamics1985} and creates self-organized ordered surface structures\cite{RosensweigInstability1967} or droplet crystals\cite{FerroCrystal2013}. A Bose-Einstein condensate with strong dipolar interactions is a quantum ferrofluid that also shows super\-flui\-dity\cite{Lahaye2007}. The field of dipolar quantum gases is motivated by the search for new phases that break continuous symmetries\cite{Lahaye2009, Baranov2012}. The simultaneous breaking of continuous symmetries like the phase invariance for the superfluid state and the translational symmetry for a crystal provides the basis of novel states of matter. However, interaction-induced crystallization in a superfluid has not been observed.
Here we use in situ imaging to directly observe the spontaneous transition from an unstructured superfluid to an ordered arrangement of droplets in an atomic dysprosium Bose-Einstein condensate\cite{DyBEC2011}. By utilizing a Feshbach resonance to control the interparticle interactions, we induce a finite-wavelength instability\cite{Santos2003} and observe discrete droplets in a triangular structure, growing with increasing atom number. We find that these states are surprisingly long-lived and measure a hysteretic behaviour, which is typical for a crystallization process and in close analogy to the Rosensweig instability. 
Our system can show both superfluidity and, as shown here, spontaneous translational symmetry breaking. The presented observations do not probe superfluidity in the structured states, but if the droplets establish a common phase via weak links, this system is a very good candidate for a supersolid ground state\cite{Andreev1969,Chester1970,Leggett1970}.
\end{abstract}
Research in condensed matter physics is driven by the discovery of novel phases of matter, in particular phases simultaneously displaying different types of order. A prime example is the supersolid state featuring simultaneously crystalline order and superfluidity\cite{Andreev1969,Chester1970,Leggett1970}. This state has been elusive and claims of its discovery in helium\cite{SupersolidHeNature,SupersolidHeScience} have been withdrawn recently\cite{AbsenceSupersolid2012}. 
One of the requirements for a spontaneous spatially ordered structure formation in the ground state of a many-body system are long-range interactions which are present in ferrofluids. As a consequence, a magnetized ferrofluid forms stationary surface waves due to a competition between gravitational or magnetic trapping, dipolar interaction and surface tension. This effect is known as the normal-field instability or Rosensweig instability\cite{RosensweigInstability1967} and leads to stable droplet patterns on a superhydrophobic surface\cite{FerroCrystal2013}. For ferrofluids the dispersion relation of surface excitations displays a minimum at finite momentum, resembling the well-studied roton spectrum in liquid helium\cite{RotonHe1961}. However, for a ferrofluid the physical interpretation of this minimum is very different than in helium. Its origin is an energy gain due to the attractive part of the dipolar interaction for clustering polarized dipoles in a head-to-tail configuration in periodic structures.
Such roton-induced structures have also been discussed for quantum ferrofluids\cite{Santos2003,AngularRoton2007}. In close similarity with a classical ferrofluid, a competition exists between the harmonic trapping, dipolar interaction and contact interaction. For increasing relative dipolar interaction, the roton instability can lead to a periodic perturbation of the atomic density distribution, closely connected to the Rosensweig instability\cite{RosensweigUeda2009}. However, it was believed that these rotonic structures would be unstable due to subsequent instabilites of the forming droplets\cite{NoSupersolid2007}.


Here we use the most magnetic element dysprosium (Dy) with a magnetic moment of $\mu = 9.93 \, \mu_\mathrm{B}$, where $\mu_B$ is the Bohr magneton, and generate a Bose-Einstein condensate (BEC)\cite{DyBEC2011}. We observe an angular roton instability\cite{AngularRoton2007, AngularCollapse2009} and find subsequent droplet formation to triangular structures with surprisingly long lifetimes. We use two key tools to study these self-organized structures. First, we use a magnetic Feshbach resonance\cite{DyChaos2015} to tune the contact interaction and to induce the droplet formation. Second, we use a microscope with high spatial resolution to detect the atomic density distribution in situ.

The first prediction of structured ground states in a dipolar BEC dates back to the early days of quantum gases\cite{DBEC2000} and the first mechanical effects have been seen with chromium atoms\cite{Stuhler2005}. There the dipolar attraction deforms the compressible gas and its shape is balanced by a repulsive contact interaction, described with the scattering length $a$. To compare the strengths of the contact and dipolar interaction, we introduce a length scale characterizing the magnetic dipole-dipole interaction strength $a_{\mathrm{dd}} = \mu_0 \mu^2 m / 12 \pi \hbar^2$. By tuning the scattering length $a$ with a Feshbach resonance such that $a < a_{\mathrm{dd}}$, the dipolar attraction dominates the repulsive contact interaction and an instability of a dipolar gas can occur\cite{Lahaye2007,Koch2008}. However, in a pancake-shaped trap the dipoles sit mainly side-by-side and predominantly repel each other due to the anisotropy of the dipole-dipole interaction, and hence the dipolar BEC is stabilized. In such a pancake-shaped configuration the roton instability at finite wavelength is predicted\cite{Santos2003,AngularRoton2007}.

\begin{figure}
	\centering
	\includegraphics[width=0.50\textwidth]{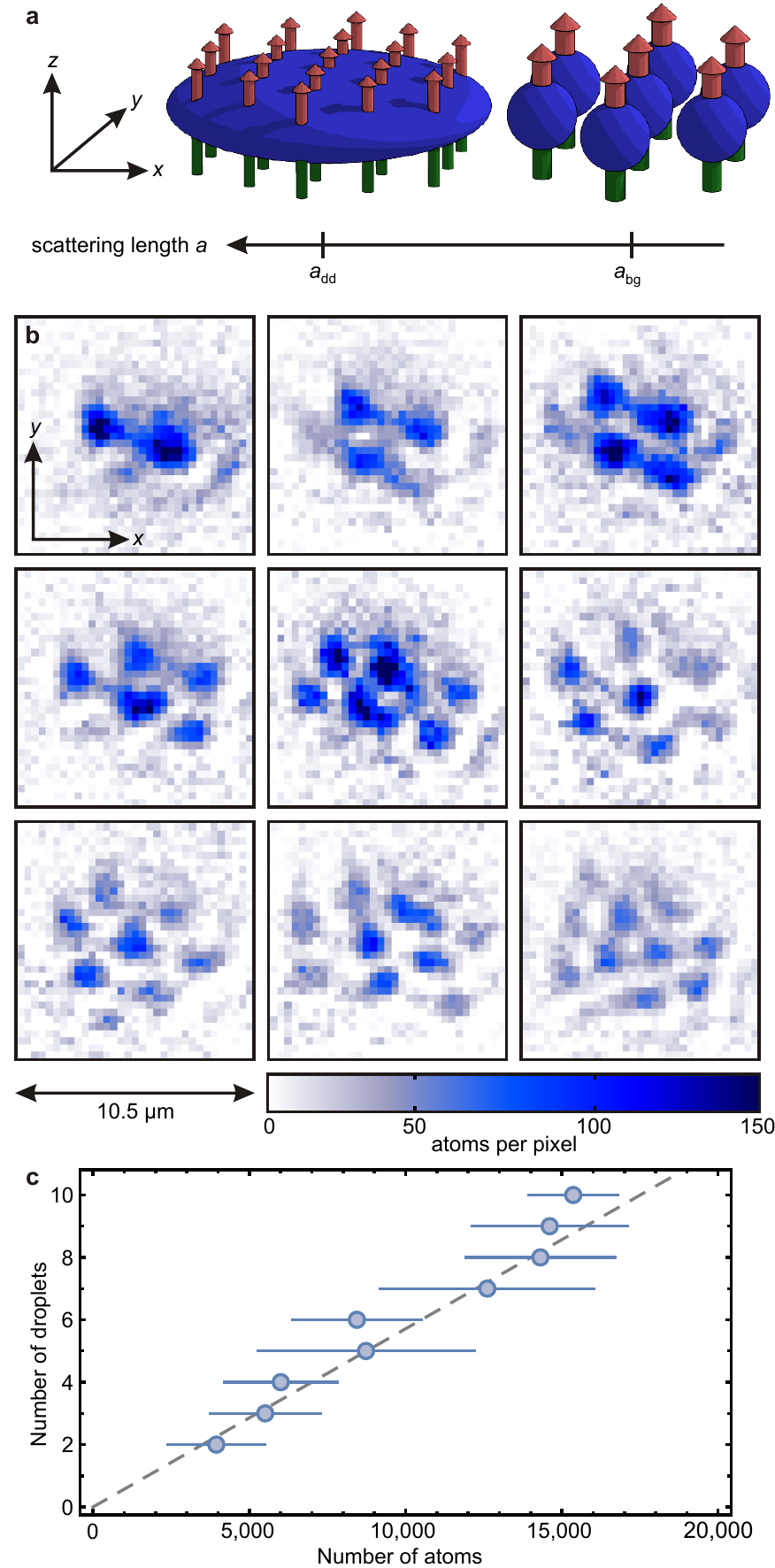}
	\caption{\textbf{Microscopic droplet crystal growth.} \textbf{a}, Scheme of the experimental procedure: We prepared a stable strongly dipolar Dy BEC at $a \approx a_{\mathrm{dd}}$ in a pancake-shaped trap. By decreasing the scattering length $a$, we induced an instability close to $a \approx a_{\mathrm{bg}}$. Followed by this instability the atoms clustered to droplets in a triangular pattern. \textbf{b}, Representative single-shot in situ images of droplet patterns with droplet numbers $N_\mathrm{d}$ ranging from two to ten. \textbf{c}, We used a set of 112 realizations with different droplet and atom numbers for a statistical analysis. The plot shows the mean number of atoms in dependence of visible droplets $N_\mathrm{d}$, with the standard deviation as error bars. We fitted a linear relation (grey dashed line) and extracted a slope of 1750(300) atoms per droplet. This shows the growth of the microscopic droplet crystal by increasing the atom number.
	\label{fig1}}
\end{figure}

For our experiments we used the isotope $^{164}$Dy with a dipolar length of $a_{\mathrm{dd}} = 132 \, a_0$, where $a_0$ is the Bohr radius. This dipolar length is already higher than the background scattering length $a_\mathrm{bg} = 92(8) \, a_0$, which is the value far from Feshbach resonances\cite{abgDy2015,BroadDy2015}. To obtain a stable BEC, we tuned the scattering length to $a \approx a_{\mathrm{dd}}$ using a magnetic field of $B = 6.962(3) \, \mathrm{G}$ in the vicinity of a Feshbach resonance located at $B_0 = 7.117(3) \, \mathrm{G}$. We then obtained typically 15,000 atoms in nearly-pure Dy BECs (see Methods section). The atoms were trapped in a radially symmetric pancake-shaped trap with harmonic frequencies of $(\nu_x,\nu_y,\nu_z) = (46,44,133) \, \mathrm{Hz}$ and the external magnetic field aligned the magnetic dipoles in axial $z$-direction. Subsequently, we tuned the magnetic field  $B \lesssim 6.9 \, \mathrm{G}$, which reduces $a$ to $~a_\mathrm{bg} < a_{\mathrm{dd}}$ resulting in an angular roton instability\cite{AngularRoton2007} that triggered the transition to ordered states (Fig.~\ref{fig1}a). We observed then the formation of droplets that arrange in ordered structures by in situ phase-contrast polarization imaging along the $z$-direction with a spatial resolution of $1 \, \mathrm{\mu m}$.
\par 
In Fig.~\ref{fig1}b we show typical in situ images of the resulting triangular patterns for the quantum ferrofluid with different number of droplets $N_\mathrm{d}$ ranging from two to ten. In order to analyze the average atom number per droplet, we count the number of droplets $N_\mathrm{d}$ in relation to the total atom number. Fig.~\ref{fig1}c indicates a linear dependence between $N_\mathrm{d}$ and atom number with a slope of 1750(300) atoms per droplet. In the case of $N_\mathrm{d}=2$ we observe a droplet distance of $d=3.0(4) \, \mathrm{\mu m}$. The droplets, having a large effective dipole moment of $N_\mathrm{d} \mu$, strongly repel each other while the radial trapping applies a restoring force. Hence, this length distance $d$ can be calculated in a simplified one-dimensional classical system by minimizing its energy. We assume two strongly dipolar particles with 1750 times the mass and magnetic moment of a dysprosium atom that are confined in a harmonic trap. For our experimental parameters they minimize their energy with a distance of $d=3.3 \, \mathrm{\mu m}$, in agreement with the observed distance. For $N_\mathrm{d}>2$ the droplets arrange mostly in a triangular structures and form a microscopic crystal with a droplet distance of $d=2-3 \, \mathrm{\mu m}$. Owing to the isotropy of the repulsion between droplets in the radial plane, we expect the triangular configuration to have the lowest energy. Because of the repelling dipolar force between the droplets, we observe in the radial direction nearly round discrete droplets with possible weak overlap to neighbouring ones.
Comparing our quantum ferrofluid with a classical ferrofluid, very similar behaviour and patterns have been observed on a superhydrophobic surface\cite{FerroCrystal2013}. In this system a single droplet first deforms for increasing external magnetic field and divides above a critical field into two droplets.
However, for a quantum ferrofluid a single droplet should be unstable for $a < a_{\mathrm{dd}}$ due to the attractive part of the dipolar interaction and collapse. In addition the counteracting quantum pressure, the zero-point energy existing due to an external trapping potential, can compensate attraction and prevent collapse\cite{LiBEC1997}. It was however predicted not to be the case by mean field calculations\cite{NoSupersolid2007, AngularCollapse2009}. Our observation of stable droplet ensembles is thus striking and further work is needed to understand their stability. A possible stabilizing effect are quantum fluctuations leading to beyond mean-field effects\cite{Lee1957}. 
Such a stabilization has been suggested in a similar situation of competing attraction and repulsion\cite{Petrov2015} and an increased effect of quantum fluctuations has been calculated for strongly dipolar gases\cite{Lima2011}.
%
\begin{figure*}
	\centering
	\includegraphics[width=0.67\textwidth]{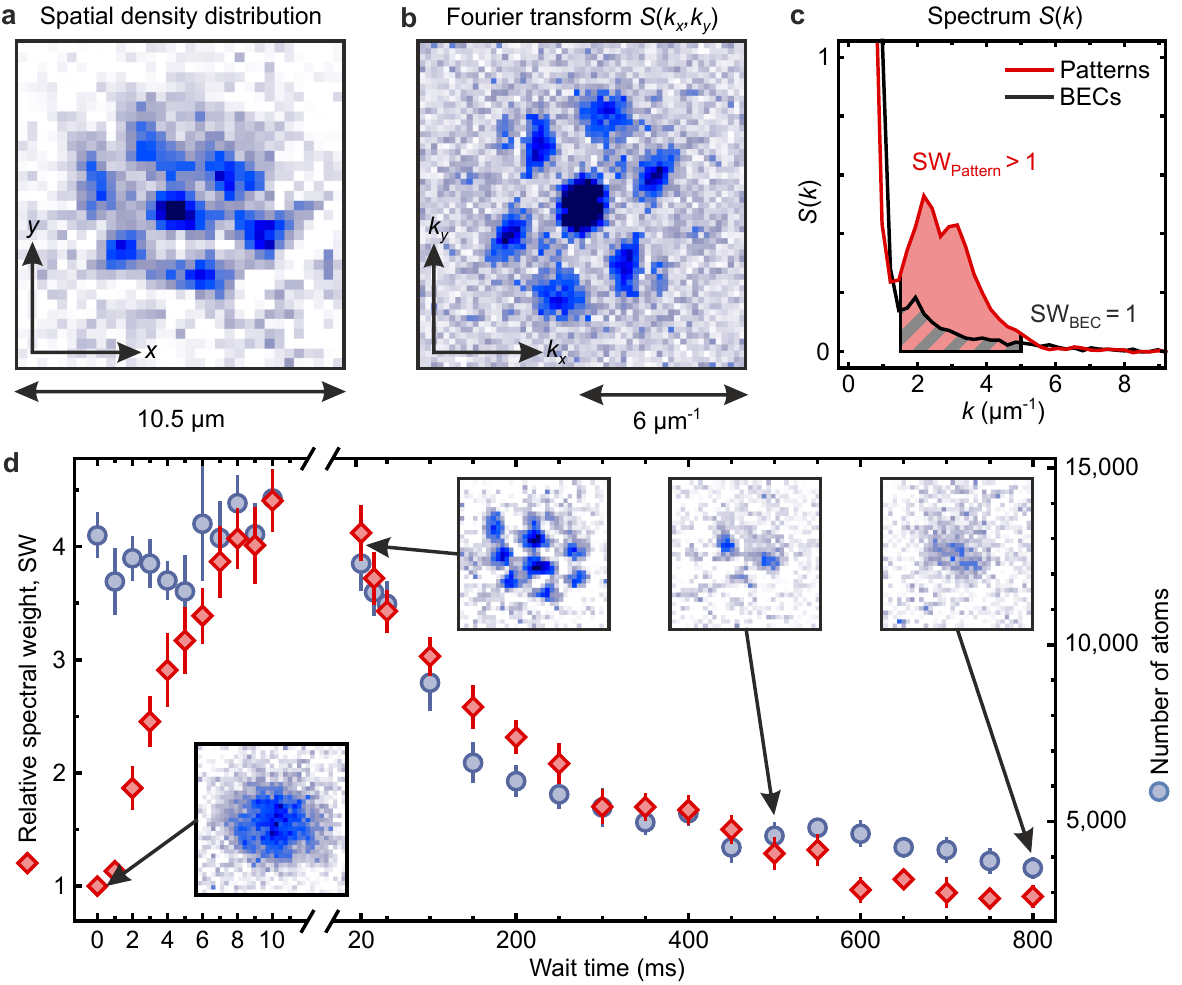}
		\caption{\textbf{Evaluation of the structures and lifetime analysis.} \textbf{a-c}, This part presents our statistical evaluation procedure. We started with \textbf{a} the spatial density distribution, calculated \textbf{b} the absolute value of the two-dimensional Fourier transform $S(k_x,k_y)$ and radially averaged over $k=(k_x^2+k_y^2)^{1/2}$ to get the spectrum $S(k)$. We removed the white noise from the spectrum $S(k)$ such that $S(k)=0$ for $k>7 \, \mathrm{\mu m^{-1}}$, which are structures below our resolution, and show in \textbf{c} an average of 13 images for BECs and Patterns. For patterns in the spatial density distribution, we observe enhanced signal for $k \approx 2.5 \, \mathrm{\mu m^{-1}}$ in the spectrum (red line), whereas $S(k)$ of BECs (black line) shows only a monotonic decay for increasing momentum. We define the sum of these spectra over a momentum range as relative spectral weight SW (coloured area, as defined in the text), which is a quantity for the strength of the structured states. \textbf{d}, We performed a sudden quench ($0.5 \, \text{ms}$) of the BEC below the instability to $B = 6.656(3) \, \mathrm{G}$ for varying wait times. To extract creation and lifetime we give the relative spectral weight SW (red diamonds) against wait time, where each point is an average of 13 realizations with error bars indicating the standard error. The plot shows a fast pattern formation within $7 \, \mathrm{ms}$ and surprisingly long 1/e-lifetimes of $\sim \!300 \, \mathrm{ms}$. The lifetime seems to be limited by losses of the atom number (blue circles). The insets are typical single-shot spatial density distributions before pattern formation (BEC) and at three different wait times.
		\label{fig2}}
\end{figure*}
\par
For further quantitative statistical analysis, we computed the Fourier spectrum $S(k)$ of the obtained images (Fig.~\ref{fig2}a-c). The patterns are visible as a local maximum at finite momentum $k=2\pi/d \approx 2.5 \, \mathrm{\mu m^{-1}}$, whereas the spectrum $S(k)$ of a BEC shows only a monotonic decrease in $k$. Hence, we define the spectral weight
\begin{equation}
	\text{SW} = \sum_{k = 1.5 \, \mathrm{\mu m^{-1}}}^{5 \, \mathrm{\mu m^{-1}}} S(k) \ ,
\end{equation}
which is a quantity for the strength of the structured states and give it in relative units such that a BEC has $\text{SW}_{\mathrm{BEC}}=1$. After a quench of the interactions from $a \approx a_{\mathrm{dd}}$ to $a \approx a_{\mathrm{bg}}$, we investigated statistically the pattern formation time and the lifetime of these patterns (Fig.~\ref{fig2}d). We repeated this measurement 13 times and found statistically that after $7 \, \mathrm{ms}$ the pattern formation is fully developed and we observed a 1/e-lifetime of $\sim \!300 \, \mathrm{ms}$. The decay of the droplet structure is accompanied by an atom loss showing a 1/e-lifetime of $\sim \!130 \, \mathrm{ms}$, while the residual thermal cloud is constant. 
Due to the decreasing atom number the structures melt back to lower number of droplets $N_{\mathrm{d}}$ until they merge back to one droplet (insets of Fig~\ref{fig2}d).
In comparison, as we measured lifetimes of a non-structured BEC of more than $5 \, \mathrm{s}$, we assume increased three-body losses as a reason for the reduced lifetime. One indication is the measured atomic peak density for droplets of $n \gtrsim 5 \cdot 10^{20} \, \mathrm{m}^{-3}$, which is larger than the density of a BEC with $n \approx 10^{20} \, \mathrm{m}^{-3}$. 
\begin{figure}
	\centering
			\includegraphics[width=0.50\textwidth]{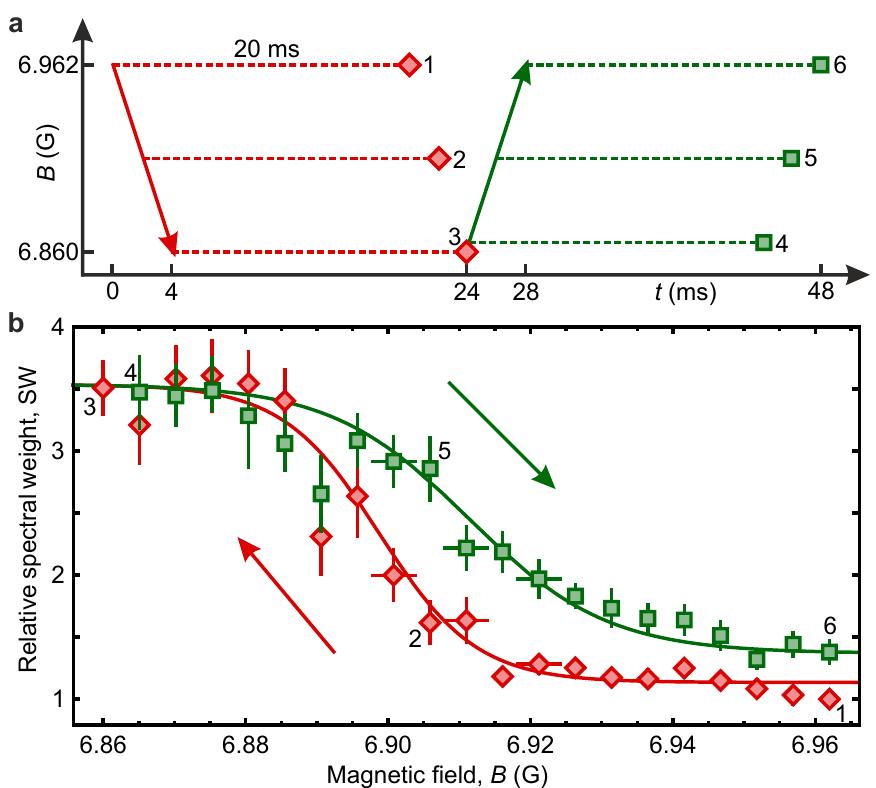}
		\caption{\textbf{Hysteresis of pattern formation.} \textbf{a}, A time line of the experiment where we observed hysteresis. We prepared the Dy BEC close to the Feshbach resonance at $6.962(3) \, \mathrm{G}$ and ramped linearly to different magnetic field values with a constant ramp speed down to a lowest value of $6.860(3) \, \mathrm{G}$. To ensure that the structures have enough time to form, we waited afterwards for $20 \, \mathrm{ms}$ and in situ imaged the atomic sample. For the way back, we waited first at the lowest field value for $20 \, \mathrm{ms}$ and then increased with the same ramp speed to different higher field values and held again for $20 \, \mathrm{ms}$. \textbf{b}, Hysteresis plot for the structured patterns. It shows the spectral weight against magnetic field for the way down (red diamonds and line) and the way back (green squares and line), each point is an average of 14 realizations with error bars indicating standard errors. Our long-term field stability was $3 \, \mathrm{mG}$ shown with horizontal error bars for selected points. A clear hysteresis is visible comparing the way down and the return, although the total wait time is twice longer for the way back. A few points are marked in \textbf{a} and \textbf{b} with numbers to help the understandability and the lines serve as a guide to the eye.
	\label{fig3}}
\end{figure}
To explore the nature of this instability, we performed the following experimental sequence, depicted in Fig.~\ref{fig3}a. We prepared the BEC close to the Feshbach resonance with $a \approx a_{dd}$ and ramped linearly to varying magnetic field values around the instability point. We ensured that the structures are formed within $10 \, \mathrm{ms}$ also for magnetic field values close to the stability threshold, and waited here for twice this time. Fig.~\ref{fig3}b shows a clear hysteresis for the way down in magnetic field compared to the return. For the way back we have the same spectral weight for $\sim \! 20 \, \mathrm{mG}$ higher magnetic field values. This demonstrates that our system features bistability in the transition region. We expect that the transition from one state to the other state is driven by thermal excitations or weak currents due to overlapping droplets. In the thermodynamic limit such behaviour is a clear signature of a first-order phase transition and a latent heat in the crystallization process.

To verify that we are dealing with a transition to a ground state and not a metastable state resulting from quench dynamics, we performed forced evaporative cooling at a constant magnetic field far away from any Feshbach resonance with $a \approx a_{\mathrm{bg}}$. We observed very similar self-organized structures, which where visible for temperatures near the expected critical temperature for the phase transition to a BEC. 


As structures can melt back into a BEC and we observed evaporative cooling to patterns, it is quite plausible that the droplets are superfluid individually. Whether they are sharing the same phase via weak links or lose their mutual phase coherence will have to be investigated in future experiments. Another open question is the creation dynamics of a self-organized structure. 
It will be interesting to study phonons in such a droplet crystal which we expect to have eigenfrequencies on the order of many inverse lifetimes. We expect that these phonons will be coupled to collective Josephson oscillations via weak links\cite{SpectrumSupersolid2012}.
	
\begin{methods}

\textbf{Preparation of a Dy BEC} 
Bosonic $^{164}\mathrm{Dy}$ atoms are first cooled in a narrow-line magneto-optical trap, operating on the $626 \, \mathrm{nm}$ optical cycling transition, which polarizes the atoms to the lowest Zeeman state $J = 8, m_J = -8,$ and subsequently directly loaded into a single-beam optical dipole trap (ODT)\cite{Maier2014} created by a broadband fiber laser operating at $1070 \, \mathrm{nm}$ with a power of $72 \, \mathrm{W}$. By moving the focusing lens of the transport ODT over a range of $375 \, \mathrm{mm}$ we transport the atoms into a glass cell. After the transport, we load the atoms from the transport ODT into a crossed optical dipole trap created by a single-mode laser operating at $1064 \, \mathrm{nm}$. Following this, the atoms are Doppler cooled with $626 \, \mathrm{nm}$ light and we utilize forced evaporative cooling by ramping down the power of both trapping laser beams at a magnetic field of $B = 1.012 \, \mathrm{G}$ far away from any resonance. When close to degeneracy we tune the magnetic field close to a Feshbach resonance and achieve there a quasi-pure BEC with typically 15,000 atoms and a temperature of $T= 70 \, \mathrm{nK}$. Before shaping the final trap with harmonic frequencies of $(\nu_x,\nu_y,\nu_z) = (46,44,133) \, \mathrm{Hz}$, we apply an additional magnetic gradient of $1.1 \, \mathrm{G/cm}$ to partially compensate gravity.

\setcounter{figure}{0}
\renewcommand{\figurename}{Extended Data Figure}

\begin{figure*}
	\centering
		\includegraphics[width=1\textwidth]{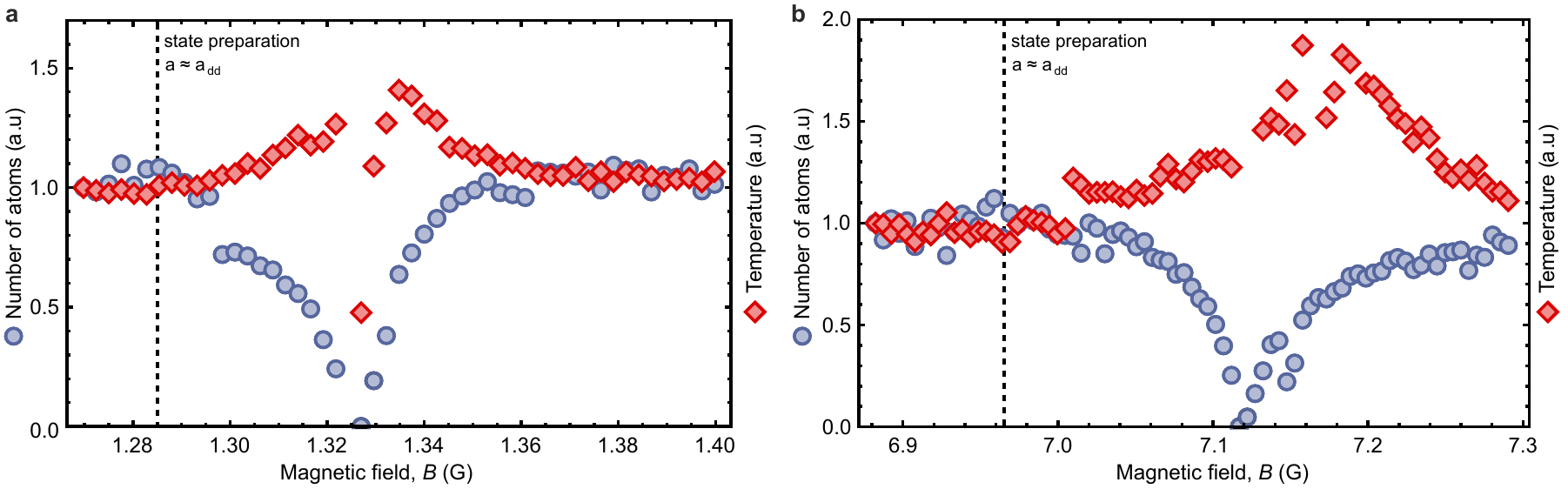}
	\caption{{\textbf{Atom trap-loss spectroscopy for two Feshbach resonances.} Atom-loss spectroscopy mapping Feshbach resonances of $^{164}$Dy. The atom number (blue circles) and temperature (red diamonds) is normalized. \textbf{a}, Atom number minimum shows the center of the Feshbach resonance at $B_0=1.326(3) \, \mathrm{G}$, while the temperature is maximized at $B_0 + \Delta B$, with $\Delta B=8(5) \, \mathrm{mG}$. We prepared stable BECs at $B=1.285(3) \, \mathrm{G}$ (dashed black line) before we induced an instability for lower magnetic field values.  \textbf{b}, We used for the investigations in this manuscript the shown resonance at $B_0 = 7.117(3) \, \mathrm{G}$ with $\Delta B = 51(15) \, \mathrm{mG}$. Stable BECs were created at $B=6.962(3) \, \mathrm{G}$ (dashed black line).}  
		\label{Extended1}}
\end{figure*}

\textbf{Feshbach resonance} 
The background scattering length of $^{164}\mathrm{Dy}$ has been measured to be $a_{\mathrm{bg}}= 92(8) \, \mathrm{a_0}$, where $a_0$ is the Bohr radius\cite{abgDy2015, BroadDy2015}. In the vicinity of a Feshbach resonance the scattering length scales as $a = a_{\mathrm{bg}} \left( 1 - \Delta B / (B-B_0) \right)$ with $B$ the applied magnetic field, $B_0$ the center and $\Delta B$ the width of the Feshbach resonance. To tune the scattering length $a$, we used a narrow resonance at $B_0 = 7.117(3) \, \mathrm{G}$ with $\Delta B = 51(15) \, \mathrm{mG}$. We calibrated the magnetic field with a weak radiofrequency field driving transitions between magnetic sublevels. To measure the resonance position we use atom trap-loss spectroscopy and find the maximal loss at the central position $B_0$ of the Feshbach resonance. To estimate their width we did a thermalization experiment at different magnetic field positions and found a maximum temperature at the field position $B_0 + \Delta B$ (Extended Data Fig.~1b). Close to the maximal temperature we observed another narrow resonance that influences our width measurement. Hence, we cannot state precise values for the scattering length. In addition, we did the same investigations for the structured states on a second even narrower resonance located at $B_0 = 1.326(3) \, \mathrm{G}$ with a width of $\Delta B = 8(5) \, \mathrm{mG}$ (Extended Data Fig.~1a). We observed the same qualitative results with this resonance, showing that self-organization is independent of a particular Feshbach resonance.

\textbf{In situ phase-contrast polarization imaging}
Phase-contrast polarization imaging was first introduced with lithium atoms\cite{LiBEC1997} and relies on the dispersive phase shift instead of direct absorption giving rise to the optical density. We use an off-resonant beam ($\Delta = -1.1 \, \mathrm{GHz} = -35 \, \Gamma$) near the strongest optical transition at $421 \, \mathrm{nm}$ with a linewidth of $\Gamma \approx 32 \, \mathrm{MHz}$ to suppress absorption. The imaging beam is linearly polarized and we apply a magnetic field of more than $1 \, \mathrm{G}$ in the beam propagation direction so that the atoms see a mixture of left- and right-handed circularly polarized light. In our case the dysprosium atoms are fully spin-polarized in the lowest lying Zeeman state $m_J=-8$. Thus the atoms couple mainly to the $\sigma^-$ optical transition, while the $\sigma^+$ transition is suppressed by a factor of $\sim \! 150$ due to the difference in the Clebsch-Gordan coefficients. Hence, the atoms show a strong circular birefringence or optical rotation. If both coupled and uncoupled polarizations are combined on a linear polarizer, the transferred intensity distribution depends on the dispersive shift of the atoms. The atomic plane is then imaged through a commercial diffraction limited objective ($\text{NA} = 0.32$), corrected for the upper window of our glass cell, with a resolution of $1 \, \mathrm{\mu m}$. With a second commercial objective the image is magnified by a factor of 50 and guided through a polarizer to the camera.

\end{methods}

\bibliographystyle{naturemag}
\bibliography{literatur}

\begin{addendum}
 \item[Acknowledgements] We would like to thank Axel Griesmaier for support at the early stage
of the experiment and we thank Damir Zajec, David Peter, Hans Peter B\"uchler and Luis Santos for discussions. This work is supported by the German Research Foundation (DFG) within SFB/TRR21. H.K. acknowledges support by the 'Studienstiftung des deutschen Volkes'.
 \item[Author Contributions] All authors discussed the results, made critical contributions to the work and contributed to the writing of the manuscript.
\end{addendum}


%
%
%

\end{document}